# Temperature dependence of the emission linewidth in MgO-based spin torque nano-oscillators

J. F. Sierra[1*], M. Quinsat[1,2], U. Ebels[1], D. Gusakova[1], I. Joumard[1], A. S. Jenkins[1], L. Buda-Prejbeanu[1], B. Dieny[1]

[1]*SPINTEC, UMR(8191) CEA / CNRS / UJF / Grenoble INP ; INAC, 17 rue des Martyrs, 38054 Grenoble, France*

M.-C. Cyrille[2]
[2]*CEA-LETI, MINATEC, DRT/LETI/DIHS, 17 Rue des Martyrs, 38054 Grenoble, France*

A. Zeltser[3] and J. A. Katine[3]
[3] *Hitachi Global Storage Technologies, San Jose, Cal. 95193, USA.*

Spin transfer driven excitations in magnetic nanostructures are characterized by a relatively large microwave emission linewidth (10 -100 MHz). Here we investigate the role of thermal fluctuations as well as of the non-linear amplitude-phase coupling parameter ν and the amplitude relaxation rate $\Gamma_p$ to explain the linewidth broadening of in-plane precession modes induced in planar nanostructures. Experiments on the linewidth broadening performed on MgO based magnetic tunnel junctions are compared to the linewidth obtained from macrospin simulations and from evaluation of the phase variance. In all cases we find that the linewidth varies linearly with temperature when the amplitude relaxation rate is of the same order as the linewidth and when the amplitude-phase coupling parameter is relatively small. The small coupling parameter ν means that the linewidth is dominated by direct phase fluctuations and not by amplitude fluctuations, explaining thus its linear dependence as a function of temperature.

[*] Corresponding author : juan.sierra@cea.fr

Self-sustained oscillations,[1] which are induced by virtue of the spin-transfer-torque effect in nanoscale ferromagnets,[2] open the possibility to conceive a new spintronics device called a spin-torque nano-oscillator (STNO). Such a nanoscale device may present a strong non-linear coupling between the oscillation amplitude and phase,[3-6] providing a way to tune the oscillation frequency by altering the current. However, theory[3-6] predicts that this coupling also produces a substantial broadening of the emission linewidth and enhances the associated phase noise, which is still too large to implement such STNOs in wireless communication applications.

One of the main sources of noise for such STNOs are thermal fluctuations of the magnetization, leading to amplitude and phase fluctuations of the magnetization trajectory.[5,7] Although the temperature dependence of the emission linewidth has been previously investigated in the frequency domain for different STNO configurations, such as metallic spin valve nanopillars[8,9] and nanocontacts[10] and magnetic tunnel junction (MTJ) nanopillars,[11] there is little quantitative analysis of the experimental linewidth in relation to the value and temperature dependence of the non-linear parameters, such as the amplitude-phase coupling $\nu$, and the amplitude relaxation rate $\Gamma_p$. The latter describes how fast perturbations of the precession amplitude are damped out. A detailed investigation of these parameters can supply additional information on the temperature dependence of the emission linewidth.

In the case of a 'linear' oscillator ($\nu=0$), it has been shown that the emission linewidth, $2\Delta f$ (defined as the full-width at half maximum (FWHM)), varies linearly with temperature.[3,4,7] However, for the case of a strong non-linear coupling ($\nu \gg 1$) the behavior can be modified depending on the relative value between $2\pi\Delta f$ and the amplitude relaxation rate $\Gamma_p$. In particular, in the high temperature limit, when $2\pi\Delta f$ is large compared to $\Gamma_p$ ($2\pi\Delta f \gg \Gamma_p$), the dependence is modified to a square-root increase with temperature.[4] In the opposite low temperature limit, when $2\pi\Delta f$ is sufficiently small compared to $\Gamma_p$ ($2\pi\Delta f \ll \Gamma_p$), theoretical predictions give a linear dependence of $2\Delta f$ with temperature, whose slope is enhanced by a factor of $(1+\nu^2)$.[4] For intermediate values, when $2\pi\Delta f$ is of the same order as $\Gamma_p$ ($2\pi\Delta f \approx \Gamma_p$), theory does not give an analytical expression for the temperature dependence and numerical evaluations are required in this case.

Here, a temperature dependent investigation is presented using frequency and time-resolved detection techniques to extract the emission frequency $f$, emission linewidth $2\Delta f$, the non-linear amplitude-phase coupling parameter $\nu$ and the amplitude relaxation rate $\Gamma_p$.[12-14] The devices investigated are MgO-based MTJ nanopillars of different diameters (85, 75 and 62 nm), *nominal* resistance per area product $RA$=1 $\Omega\mu m^2$ and with tunnelling magnetoresistance (TMR) values around 85% at room temperature.

The samples have been fabricated by sputter deposition with the following composition: IrMn(6.1) / CoFe(1.8) / Ru / CoFeB(2) / MgO(0.9) / CoFe(0.5) / CoFeB(3.4) (thicknesses in nm). While precessional dynamics is induced in the CoFe / CoFeB free layer, the CoFe / Ru / CoFeB synthetic antiferrromagnet (SAF), which is exchange biased by the IrMn layer, acts as the polarizing layer. Measurements were performed for in-plane magnetic fields $H_{app}$ applied close to the easy-axis magnetization of the free layer (offset angle ~10°). We used a cryogenic probe station which allows measurements with temperature stabilization better than 100 mK. The temperature quoted here corresponds to the one of the thermal bath and not to the device temperature. If not specified otherwise, the data correspond to an 85 nm diameter nanopillar device (DevA) for which the *measured RA* value was 1.4 $\Omega\mu m^2$ and the TMR varies linearly from 86% at room temperature to 105% at 100 K. A similar behaviour for the temperature dependence of the emission linewidth has been observed for about ten devices.

Dynamic excitations for the free layer have been measured when the magnetization of the free and the top SAF layer were aligned antiparallel (positive fields) and when electrons are flowing from the polarizing to the free layer (negative bias currents). The excitation spectrum is characterized by a single microwave emission mode (see inset Fig. 1(a)). The dependence of its frequency $f$ on the in-plane field shows a Kittel-like increase with several branches (Fig. 1(a)). The latter are separated by small frequency jumps which lead to an increase in linewidth (Fig. 1(b)) due to telegraph noise like transitions between the two branches. In consequence the linewidth is minimum only in the center of each branch. This trend has been previously observed in similar devices at room temperature.[15] For the devices studied here this branch structure is temperature independent in the range of temperatures covered experimentally (100 K-300 K).

In order to analyze the temperature dependence of the linewidth we adjusted the in-plane field to be in the center of one of the branches and far from the frequency jumps. This choice eliminates contributions to the linewidth broadening arising from stochastic transitions and is thus closer to the intrinsic behavior. For such fields we have measured the evolution of $f$ and $2\Delta f$ with bias current. For DevA, shown in Figs. 1(c) and 1(d), this field is $\mu_0 H_{app}$ = 61 mT. The critical current, $I_C$, above which self-sustained oscillations are excited, has been determined from the amplitude histograms extracted from the time traces of the oscillating voltage signal as shown in Ref. 16. For the given field value we deduced a value of $I_C$= -0.80 mA, which is independent of temperature. This temperature independence can be explained by small changes of the spin current polarization $P$ ($P_{100K}/P_{300K}$ = 1.05, extracted from the TMR *vs.* temperature) that are compensated by small variations of the saturation magnetization $M_S$ ($M_{S(100K)}/M_{S(300K)}$ = 1.08).[1] The value of $M_S$ was deduced from the thermally induced ferromagnetic resonance mode (T-FMR) in the subcritical region $I<I_C$, where $f$ slightly increases with decreasing temperature.

The central result of this investigation is presented in Fig. 2(a), where the temperature dependence of the linewidth is shown for three different devices (DevA = 85 nm, DevB = 75 nm and DevC = 62 nm diameter) for a bias current $I>I_C$. Two important features are common to all devices. Firstly, the linewidth increases linearly with temperature with an average slope of 0.25 MHz/K and its value changes by about a factor of two between 100K and 300K. For instance, for device DevA, the linewidth increases from 25 to 50 MHz. Secondly, the interpolation of the linewidth from 100 K to 0 K shows a positive intercept. Measurements below 100 K (down to 5 K) present a saturation of the emission linewidth (i.e. an almost constant value) that is accompanied by a saturation of the emission frequency and the device resistance. From this we conclude that Joule heating saturates the device temperature and thus it is not possible to investigate this temperature range with good confidence. Subtracting the zero intercept and plotting the linewidth as a function of temperature on a log-log scale (not shown here) gives a slope $2\Delta f$ vs. temperature equal to one for all the devices thus confirming the linear dependence.

In order to elucidate the role of the non-linear parameters $\nu$ and $\Gamma_p$ to explain the linear temperature dependence of the linewidth we have extracted these parameters from the voltage time traces applying the Hilbert transform formalism[12,13] which provides the phase $\Phi$ and the power fluctuations $\delta p$. Evaluation of the autocorrelation function of $\delta p$ shows an exponential decay which allows the extraction of the amplitude fluctuation correlation time $\tau=1/2\Gamma_p$. Figure 2(b) shows the temperature dependence of both $\Gamma_p$ and the relative ratio $r=2\pi\Delta f/\Gamma_p$. The data show that $\Gamma_p$ increases slightly with temperature, while $r$ is close to, but larger than one for all temperatures. This indicates that neither the high temperature limit ($r>>1$) nor the low temperature limit ($r<<1$) is applicable to our samples to explain the temperature dependence of the linewidth. We therefore have extracted the power spectral densities of the amplitude and phase noise, $S_{\delta p}$ and $S_\Phi$ respectively (see Fig. 2(c)), to estimate the non-linear phase-amplitude coupling parameter, $\nu$ (see Ref. 12 for details). Theory shows that $S_\Phi$ has a linear and a non-linear contribution,[5,6] the former being due to direct noise on the phase and the latter arising from amplitude noise via the non-linear amplitude-phase coupling. From the ratio between this linear and non-linear contribution[12], we obtain a value of ν<2 for all temperatures (see Fig. 2(c)) for the different devices.

Since the linewidth is of the order of $\Gamma_p$ and since ν is small, we have performed numerical calculations to check for the expected temperature dependence of the linewidth for these parameters. This has been done in two different ways: (i) using numerical simulations for self-sustained IPP modes and (ii) evaluating the linewidth via the phase variance using the experimental non-linear parameters $\Gamma_p$ and ν as input parameters.[4,5]

For the numerical simulations, case (i), the Landau-Lifshitz-Gilbert-Slonczewski equation, with an additional white Gaussian thermal noise field, was solved in the macrospin approach.[17] The parameters used in the simulations were close to the experimental parameters of the free layer: thickness $t$ = 3.9 nm, $M_S$ = 1000 kA/m, damping constant $\alpha$ = 0.02, shape anisotropy field $\mu_0 H_K$ = 8.614 mT, in-plane field $\mu_0 H_{app}$ = 40 mT and $I/I_C$ = 1.6. We note that we adapted the length of the simulated time trace such that the frequency resolution remained smaller than the linewidth (in particular at low temperatures). The results are shown in Fig. 2(d)–(e), revealing a linewidth that varies linearly with temperature in the range 1 K<T<300 K, with a slope of 0.46 MHz/K. In contrast to the experimental data, the extraction of $\Gamma_p$ gives a constant value ($\Gamma_p/\pi \approx$ 250 MHz) in the range 100 K<T<300 K and a relative ratio $r$ close to, but smaller than one (see Fig. 2(e)). The values are similar to the experiments and confirm that for the IPP mode with fields applied close to the easy-axis, neither the high nor the low temperature limit of the theoretical model can be applied directly. The extracted amplitude-phase coupling gives a value of ν =3.7-4.02 for all the temperatures. Although this value of ν is larger than the experimental one, we emphasize that the temperature dependence of the linewidth is linear from 1 to 300 K, confirming the experimental results.

As pointed out above, the amplitude relaxation rate $\Gamma_p$ in the experiment changes with temperature while in the simulation it is constant. The origin of this temperature dependence in the experiment is not clear, but might be related to the changes in the material parameters. Note, in the simulation the materials parameters did not change as a function of temperature. To verify, whether this temperature dependence of $\Gamma_p$ can change the linear temperature dependence of 2Δ$f$, we have evaluated the linewidth from the phase variance ΔΦ² (case ii) following the description in ref. 4. As input parameters to calculate ΔΦ² we have used the experimental values of $\Gamma_p$(T) given in Fig. 2(b) and of ν=2 . Furthermore, we have used for the linear linewidth $2\Delta f_0/T$ a value of 0.1MHz/K.[12] The result is shown in the inset of Fig. 2(d) in a log-log plot, revealing a slope of one for $2\Delta f$ vs. temperature. This means that the temperature dependence of $\Gamma_p$ does not alter the overall linear linewidth dependence.

While both the numerical simulations and the evaluation of the phase variance confirm the linear dependence of $2\Delta f$ vs. temperature found in the experiments for the given values of $\nu$ and $\Gamma$, there is one substantial difference. From theory the linewidth is expected to go to zero at 0K, while the experimental results show a non-zero intercept for the linewidth for all devices (compare Figs. 2 (a) and (d)). The corresponding values for the zero-temperature linewidth offset are $2\Delta f(T=0K)=10-50$ MHz. It is noted that a similar offset in linewidth has been reported previously for spin valve nanocontacts which however showed positive and negative offset values. In contrast, for the IPP modes excited in the MTJ nanopillars investigated here, we only observe positive offsets. Due to the small value of $\nu$ we exclude that this positive offset is due to a non-linear temperature dependence of the linewidth, like a square root dependence predicted by theory for $2\pi\cdot\Delta f/\Gamma_p \gg 1$. We rather attribute the finite offset in linewidth to either a shift in temperature or a possible temperature independent broadening mechanism.[10,11] The shift in temperature may arise from the difference between the thermal bath temperature and the device temperature (larger than the bath temperature). A temperature independent contribution might arise for instance from noise in the bias current or in the tunnel current, from fluctuations in the polarizer magnetization or from the preponderance of chaotic dynamics in the free layer magnetization.[10,11]

Before closing we would like to make three further comments.

(1) The above mentioned experimental results are *only* valid for in-plane fields far from the frequency jumps observed in the $f$ vs. $H_{app}$ curves, see Fig. 1(a) and (b). For in-plane fields close to the frequency jumps (where $2\Delta f$ shows a relative maximum) the dependence of the emission linewidth on the temperature shows an exponential decrease with decreasing temperatures (see Fig. 3(a)). As mentioned above, this is attributed to a stochastic transition between the branches similar to results previously reported for spin valve nanopillars.[9]

(2) In the subcritical regime $I \ll I_C$, where the T-FMR mode is detected, the FMR linewidth is almost flat down to 150 K and increases strongly below this temperature (see Fig. 3(b)). This increase has been observed in continuous film AlO-based MTJ multilayers[18] and MgO-based MTJs nanopillars[11] and could be related with oxygen contents in the free layer[19] or a temperature induced interface anisotropy in the ferromagnet/insulator interface.[18]

(3) The results presented here were performed on devices that we called H(igh)-TMR in our previous studies.[15] These HTMR devices are characterized by a relatively homogeneous tunnel barrier. We also performed temperature dependent studies on devices called L(ow)-TMR, which are characterized by the presence of pinholes in the tunnel barrier leading to a different mode character of the excitations.[15] We found that the linewidth of the LTMR devices does not change upon decreasing temperature, taking care that the field was adjusted so as to be in the center of a branch, compare Fig. 1(a). This indicates that in the LTMR devices either the linewidth is given by a temperature independent broadening mechanism or by a strong local heating due to the pinholes that lead to a temperature of the excitation volume that does not change with bath temperature. The fact that the room temperature linewidth in LTMR devices (10-40 MHz) is less than the corresponding linewidth of the HTMR devices (50-100MHz) but of the order of the zero temperature offset linewidth, we conclude that a temperature independent broadening mechanism is the dominant contribution in the LTMR devices.

In summary, the temperature dependence of the emission linewidth of self-sustained oscillations in MgO-based magnetic tunnel junction nanopillars has been investigated experimentally by means of frequency and time resolved detection techniques, which allow the extraction of the emission frequency, emission linewidth and the non linear parameters. The results are compared to the linewidth obtained from macrospin simulations and from the phase variance using the experimental parameters $\nu$ and $\Gamma_p$. In all cases, a linear variation of the emission linewidth with temperature is observed. Since the ratio $2\pi \cdot \Delta f / \Gamma_p$ is close to one in the experiment and macrospin simulations, neither the low temperature nor the high temperature limit predicted by theory is applicable. The linear dependence is therefore understood as a consequence of the weak amplitude-phase coupling of the in-plane precession mode excited in our devices. It would be of interest to perform a similar analysis as presented here for modes with larger values of $\nu$, for instance for the perpendicular polarizer structure.[20] In this case out-of-plane precession modes are induced, that are characterized by large values of $\nu \approx 10$–$50$.[21] One could expect that in this case the role of the amplitude-phase coupling on the temperature dependence of the emission linewidth is more pronounced.

A further consequence of the weak amplitude-phase coupling in the case of the IPP mode is that the linewidth broadening is dominated by the thermally driven phase noise acting directly on the phase and not by the non-linear amplitude phase coupling. Therefore in this case, reducing phase noise and linewidth for technological applications at room temperature should concentrate on reducing the slope of 2$\Delta f$ vs. temperature.

The authors would like to thank W. E. Bailey, F. Garcia-Sanchez, V. Tiberkevich and A. Slavin for fruitful discussions. J.F. Sierra acknowledges support from the FP7-People-2009-IEF Program No 252067. All the authors acknowledge support from ANR-09-NANO-037, Carnot-RF and Nano2012 convention projects.

# Figures

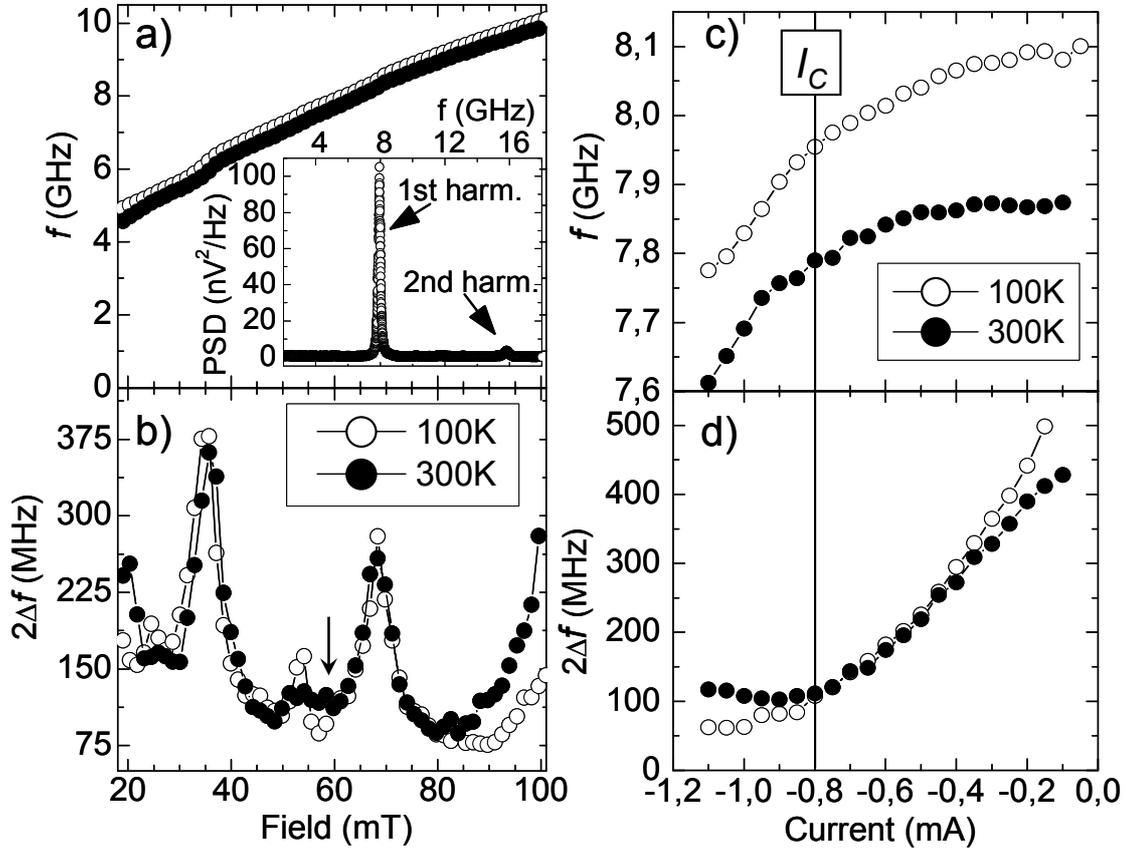

**Fig 1.** (a) Emission frequency $f$ as a function of the in-plane field for $I= -0.85$ mA. Inset: Power spectral density (PSD) at $\mu_0 H_{app}= 61$ mT. (b) FWHM emission linewidth $2\Delta f$ as a function of the in-plane bias field for $I= -0.85$ mA. Four different branches are clearly distinguished. The arrow indicates the field chosen for DevA. (c) Emission frequency $f$ as a function of the bias current and (d) FWHM emission linewidth $2\Delta f$ as a function of the bias current at $\mu_0 H_{app}= 61$ mT. In the four graphs data are shown at 100 K (open symbols) and 300 K (full symbols).

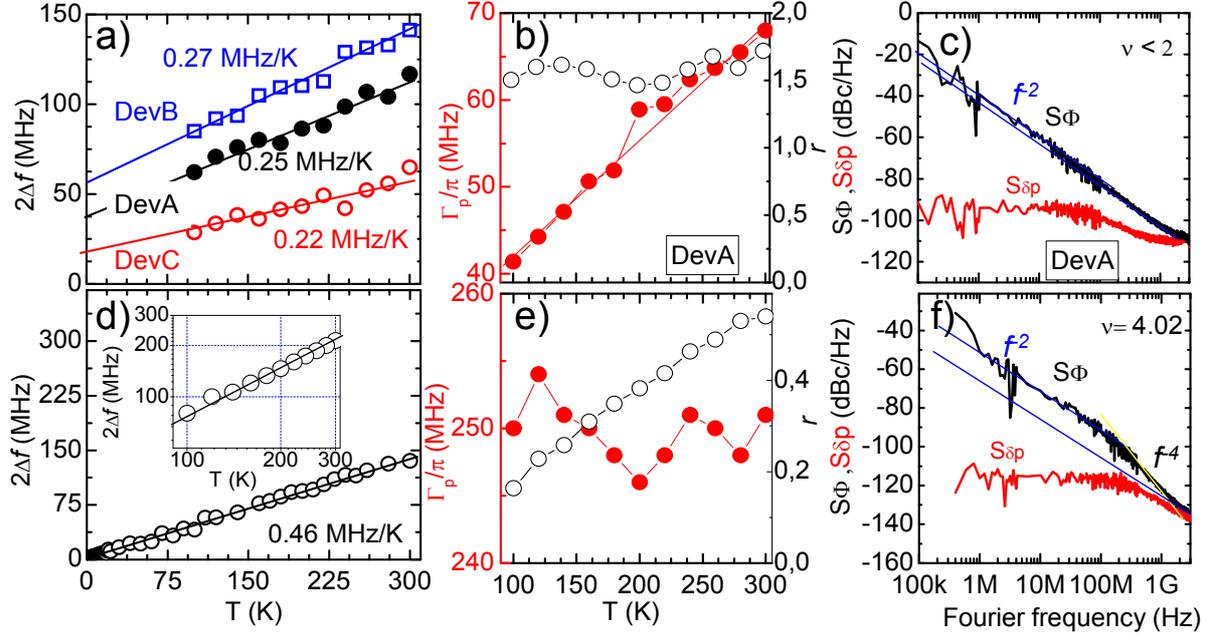

**Fig 2.** Top panel (a) – (c) (Experimental results): Temperature dependence of: (a) the FWHM emission linewidth $2\Delta f$ for three different devices: DevA: 85 nm diameter ($\mu_0 H_{app}$= 61 mT; $I/I_C \approx 1.4$), DevB: 75 nm ($\mu_0 H_{app}$= 48 mT; $I/I_C \approx 1.3$) and DevC: 62 nm ($\mu_0 H_{app}$= 65 mT; $I/I_C \approx 1.4$). (b) The amplitude relaxation rate $\Gamma_p$ (left axis) and the relative ratio $r=2\pi\Delta f/\Gamma_p$ for a bias current of -1.10 mA, i.e, $I/I_C \approx 1.4$, (right axis) for DevA. (c) Log-log plot of the phase noise $S_\Phi$ and amplitude noise $S_{\delta p}$ at 300K for the DevA. The extracted $\nu$ is given in the figure. Bottom panel (d)-(f) (macrospin simulations): Temperature dependence of (a) the FWHM emission linewidth $2\Delta f$. Inset: For comparison, $2\Delta f$ extracted via the phase variance (see text), (d) the amplitude relaxation rate $\Gamma_p$ (left axis) and the relative ratio $r=2\pi\Delta f/\Delta\Gamma_p$ (right axis). (f) Log-log plot of the phase noise $S_\Phi$ and amplitude noise $S_{\delta p}$ at 300K. The extracted $\nu$ is given in the figure.

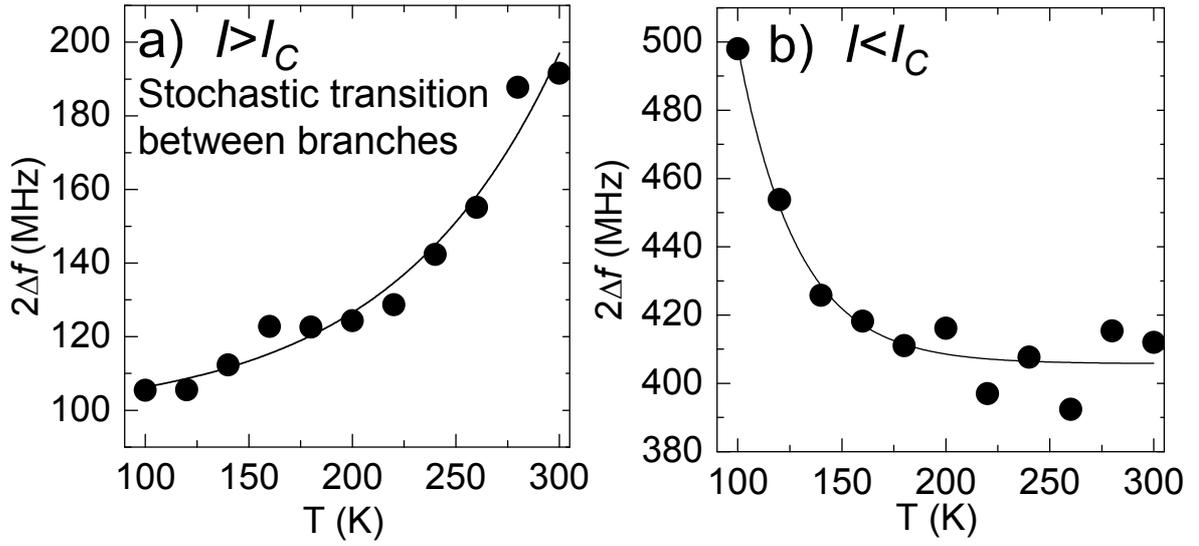

**Fig 3**. Dependence of the FWHM linewidth $2\Delta f$ on the temperature (a) for $I>I_c$ and an in-plane field of $\mu_0 H_{app}$=53.5 mT, which is close to the frequency jump (see Fig. 1(a)). (b) for $I<I_c$, *i.e*, for the T-FMR mode at $\mu_0 H_{app}$= 61 mT.